\documentclass[10pt,journal,compsoc]{IEEEtran}
\pdfoutput=1

\usepackage[bindingoffset=0.2in,%
            left=0.45in,right=0.58in,top=1in,bottom=1in,%
            footskip=.25in]{geometry}
\usepackage{graphicx}
\usepackage{cite}
\usepackage{amsmath}
\usepackage{amssymb}
\usepackage{epstopdf}
\usepackage{algpseudocode}
\usepackage{array}
\usepackage{balance}
\usepackage{multirow}
\usepackage{bigstrut}
\usepackage{tabularx}
\usepackage[table,xcdraw]{xcolor}
\usepackage[super]{nth}
\usepackage{subfig}
\usepackage{filecontents}
\usepackage[colorlinks, citecolor=blue, urlcolor=black]{hyperref}
\usepackage[square,sort,comma,numbers]{natbib}

\IEEEoverridecommandlockouts
\begin{document}

\newcommand\MYhyperrefoptions{bookmarks=true,bookmarksnumbered=true,
pdfpagemode={UseOutlines},plainpages=false,pdfpagelabels=true,
colorlinks=true,linkcolor={black},citecolor={black},urlcolor={black},
pdftitle={Bare Demo of IEEEtran.cls for Computer Society Journals},
pdfsubject={Typesetting},
pdfauthor={Michael D. Shell},
pdfkeywords={Computer Society, IEEEtran, journal, LaTeX, paper,
             template}}


\hyphenation{op-tical net-works semi-conduc-tor}


\title{Sherlock in OSS: A Novel Approach of Content-Based Searching in Object Storage System 
}

\author{
\IEEEauthorblockN{
Jannatun Noor, 
Rizwanul Haque Ratul*,
Mir Rownak Ali Uday*,
Joyanta Jyoti Mondal*, \\
Md. Sadiqul Islam Sakif*\thanks{*These authors contributed equally to this work.},
A. B. M. Alim Al Islam
\thanks{ • Jannatun Noor is with the Department of CSE, Bangladesh University of Engineering and Technology, Dhaka, Bangladesh, and with Computing for Sustainability and Social Good (C2SG) Research Group, School of Data and Sciences, BRAC University, Dhaka, Bangladesh. Email: jannatun.noor@bracu.ac.bd}
\thanks{ • Rizwanul Haque Ratul is with C2SG Research Group, and Optimizely, Dhaka, Bangladesh. Email: rizwanulratul192@gmail.com
}
\thanks{ • Mir Rownak Ali Uday is with C2SG Research Group. Email: mir.rownak.ali.uday@g.bracu.ac.bd}
\thanks{ • Joyanta Jyoti Mondal is with the Department of Computer Science, University of Alabama at Birmingham, United States, and with C2SG Research Group. Email: jmondal@uab.edu}
\thanks{ • Md. Sadiqul Islam Sakif is with C2SG Research Group, and with mPower-Social Enterprise, Dhaka, Bangladesh. Email: md.sadiqul.islam.sakif@g.bracu.ac.bd }
\thanks{ • A. B. M. Alim Al Islam is with Next-generation Computing Research Group, Department of CSE, Bangladesh University of Engineering and Technology, Dhaka, Bangladesh. Email: alim\_razi@cse.buet.ac.bd}
{}
{}
{}
}


}

\markboth{Journal of IEEE Transactions on Parallel and Distributed Systems}%
{Shell \MakeLowercase{\textit{et al.}}: Bare Advanced Demo of IEEEtran.cls for IEEE Computer Society Journals}
%



\IEEEtitleabstractindextext{%

\begin{abstract}
Object Storage Systems (OSS) inside a cloud promise scalability, durability, availability, and concurrency. However, open-source OSS does not have a specific approach to letting users and administrators search based on the data, which is contained inside the object storage, without involving the entire cloud infrastructure. Therefore, in this paper, we propose Sherlock, a novel Content-Based Searching (CoBS) architecture to extract additional information from images and documents. Here, we store the additional information in an Elasticsearch-enabled database, which helps us to search for our desired data based on its contents. This approach works in two sequential stages. First, the data will be uploaded to a classifier that will determine the data type and send it to the specific model for the data. Here, the images that are being uploaded are sent to our trained model for object detection, and the documents are sent for keyword extraction. Next, the extracted information is sent to Elasticsearch, which enables searching based on the contents. Because the precision of the models is so fundamental to the search's correctness, we train our models with comprehensive datasets (Microsoft COCO Dataset for multimedia data and SemEval2017 Dataset for document data). Furthermore, we put our designed architecture to the test with a real-world implementation of an open-source OSS called OpenStack Swift. We upload images into the dataset of our implementation in various segments to find out the efficacy of our proposed model in real-life Swift object storage.
\end{abstract}


}

\maketitle

\IEEEdisplaynontitleabstractindextext

%
\IEEEpeerreviewmaketitle

\ifCLASSOPTIONcompsoc
\IEEEraisesectionheading{\section{Introduction}\label{sec:introduction}}
\else
\vspace{-.7cm}
\section{Introduction}
\label{sec:introduction}
\fi

\vspace{-.2cm}

\IEEEPARstart{T}remendous amount of data is being produced every day which is stored and retrieved from various cloud servers. A recent estimation \cite{Duarte2023-gv} shows people create about 328.77 million terabytes of data on a daily basis. In order to store huge amounts of data, object storage is well-known to have an upper hand because of its flexibility and consistency. One of the distributed and consistent open-source cloud systems is OpenStack. OpenStack provides cloud object storage, allowing us to preserve and pull up large amounts of data using an API called OpenStack Swift. OpenStack Swift is scalable and has been designed to be durable, and available for the whole data set. Swift is a well-suited storage system for unstructured data that can be increased immensely \cite{noor2021strategizing}. Swift object storage stores every single piece of data as an object, unlike the storage systems, for instance, file-based storage or block storage, which stores data as a file. This storage system is built to house massive amounts of data at a time because of its flexibility. The retrieval of relevant data has become a significant issue as the amount of data increases significantly \cite{intro-5}. Storing consumer and business data in either public clouds or private clouds has made it difficult to efficiently and effectively retrieve meaningful data \cite{intro-5}. 

As a result, cloud-based storage is being developed using object storage. Various well-known cloud service providers such as Amazon S3, OpenStack Swift, Caringo Swarm, and many others provide object storage. Although the problem of storing massive amounts of structured and unstructured data is solved due to the complex architecture of object-based storage systems, retrieving or searching for a certain object/file has become a major challenge \cite{ref-23}. Object storage, for instance, OpenStack Swift uses the HEAD or GET method in order to get an object from the storage which is not very efficient when it comes to content-based searching. Moreover, the exact path of the object is also needed in order to get some data from this storage system, which is an inefficient task if there are massive amounts of data.

Even so, compared to block and file storage, object storage may produce higher delay and require more processing time, but it also has a number of advantages, including scalability, cost-effectiveness, robustness, and easier management. It offers great redundancy and data durability, making it particularly useful for managing massive amounts of unstructured data. Another benefit of using Open Stack Swift as an Object Storage is, Swift object storage allows simultaneous access from several servers, therefore server binding is not a problem. 
It provides fault tolerance, scalability, and adaptability without impairing system performance.

Furthermore, the linear searching method inside this storage is very time-consuming as the different replica copies are located in different regions. Searching in object storage is not significant or efficient that way, since it stands.
Content-Based Search (CoBS) primarily denotes the search that investigates the contents of inputted data rather than the metadata connected with the data, such as keywords, tags, or descriptions. In this usage, ``content'' may indicate colors, forms, materialistic details, or any other information obtained from the data itself. Manually annotating photos by inserting keywords or information into a huge database takes time and may not catch the keywords intended to identify the data. 

The interest in CoBS is starting to grow as the usage of data is increasing and metadata-based systems are struggling to work on a large amount of data. Existing technology can rapidly search for information about any data, but this requires humans to manually characterize each image in the database. This can be difficult for extremely big databases or photos created automatically, such as those from surveillance cameras. Images that utilize various synonyms in their descriptions may also be overlooked. Systems based on classifying photos in semantic classes like ``cat'' as a subclass of ``animal'' may avoid the miscategorization problem, but will take more labor by a user to locate images that may be ``cat'', but are only classed as an ``animal''. Many standards for categorizing photos have been proposed, but all encounter scaling and miscategorization difficulties.
Besides, to our knowledge, there has been no work based on content searching across different types of information in one architecture.

In this paper, we propose Sherlock, a CoBS architecture for an object storage system that enables us to extract additional metadata from images and keywords from documents and store them in a metadata database that helps us search for our desired data based on its contents. 
In our paper, we are referring to content as the objects present in images and documents. In order to do so, we first identify the type of the file. Considering the file is an image, we extract the information using an object detection Convolutional Neural Network (CNN) model named DarkNet, YOLOv4\cite{ref-36} and YOLOv8\cite{jocher2023yolo} architecture to detect objects. On the other hand, for document files, we extract the information using one of the Natural Language Processing (NLP) architectures, BERT \cite{Devlin2018-sw}. Afterwards, we retrieve additional data such as the object path in the form of an HTTP link. The data is passed to an Elasticsearch Cluster (ESC)\cite{ref-44} and the object is uploaded to object storage systems like OpenStack Swift. When the user searches for an object, our proposed interface takes input from the user, performs a search in the ESC, and returns a list of objects. The user can be able to access the objects from Swift. In this way, there is only one GET request to the object storage system. Besides, the enriched content metadata are created using BERT and DarkNet and stored in ESC ensuring more relevant content searching for the user. 

 
We propose the following contributions to this paper based on our findings -
\begin{itemize}
\item Our work is the first to come up with an architecture that can jointly do CoBS inside images and documents.
\item We create the OpenStack Swift JOSS client User Interface (UI) in order to access Swift and the Elasticsearch cluster at the same time using user-level authentication tokens.
\item We rigorously test our BERT, YOLOv4 (Darknet), and YOLOv8 algorithms with different custom-weighted files to get maximum accuracy. We use three different datasets and calculate the response time of the YOLOv4 and YOLOv8 object detectors as well as their precision in detecting multiple objects in a single image.
\item We use a pre-trained BERT model to extract keywords from Documents. Besides, in the Elasticsearch cluster, we perform multiple query requests from our User Interface to get the elastic cluster's response time and the average query time. Lastly, we add different filters to the search engine using the elastic cloud API.
\end{itemize}




\section{Related Work}\label{related_work}

The concept of searching in Object storage is not entirely new to us. Platforms like Amazon S3, and OpenStack Swift- all have their own kind of searching approaches. Although there has been very little research related to searching in object storages they have not been implemented on platforms like OpenStack Swift and other object storages. As a result, we study these research papers in order to understand their work and the complications they faced while working with object storage. We segment the search into two different categories. The first part aims to review previous relevant works in the field of searching in object storage, specifically different types of metadata-related searches. The second part emphasizes on Query related searches in Object storage.

\subsection{Metadata Searching}
Leung et al. \cite{lit-1} suggest a scalable index-based file metadata search system that outperforms competing solutions in terms of query performance while using less disk space, \cite{lit-1} named ``The SpyGlass''. The type of programs that operate with millions of data generate need to be analyzed petabytes of data which are divided into millions of files or objects, according to Singh et al. \cite{lit-2}. 
They propose a Metadata Catalog Service (MCS) which can store and access different types of metadata, and users can query for any type of metadata they want.
 


\begin{table*}[]
\centering
\resizebox{\textwidth}{!}{%
\begin{tabular}{|l|l|l|l|}
\hline
\multicolumn{1}{|c|}{\textbf{Reference}} &
  \multicolumn{1}{c|}{\textbf{Purpose of the study}} &
  \multicolumn{1}{c|}{\textbf{Technologies used}} &
  \multicolumn{1}{c|}{\textbf{Findings}} \\ \hline
Ren et al., 2023.\cite{Ren_2023} &
  \begin{tabular}[c]{@{}l@{}}Image Retrieving from large\\  amount of image pool\end{tabular} &
  Zero-shot, NormSoftmax loss, &
  \begin{tabular}[c]{@{}l@{}}Challenges and solution \\ for image retrieving \\ from sketch-based images\end{tabular} \\ \hline
Kakizaki et al., 2023.\cite{10030859} &
  \begin{tabular}[c]{@{}l@{}}Defense approach for deep \\ neural network (DNN) based \\ content-based image retrieval (CBIR)\end{tabular} &
  Interval bound propagation (IBP), DNN &
  \begin{tabular}[c]{@{}l@{}}Attack defense against \\ content-based image \\ retrieving\end{tabular} \\ \hline
Wang et al., 2023.\cite{WANG2023103163} &
  \begin{tabular}[c]{@{}l@{}}Content based image \\ retrieving in industrial section\end{tabular} &
  CNN, Binarization, Pre-trained AlexNet &
  \begin{tabular}[c]{@{}l@{}}Deep learning to retrieve \\ image with feature \\ extraction and use of \\ binarization in CBIR\end{tabular} \\ \hline
Choe et al., 2022.\cite{Choe2022-as} &
  \begin{tabular}[c]{@{}l@{}}Deep learning based \\ image retrieving\end{tabular} &
  CNN &
  \begin{tabular}[c]{@{}l@{}}CNN based image \\ retrieving improved \\ accuracy than before\end{tabular} \\ \hline
Monowar et al., 2022.\cite{Monowar2022-eg} &
  \begin{tabular}[c]{@{}l@{}}New AutoRet content-\\ based image retrieval system\end{tabular} &
  Deep CNN, AutoEmbedder framework &
  \begin{tabular}[c]{@{}l@{}}Functionality of a CBIR \\ system with spatial pooling, \\ DCNN\end{tabular} \\ \hline
Keisham et al., 2022.\cite{Keisham2022-ip} &
  \begin{tabular}[c]{@{}l@{}}Search and Rescue based \\ CBIR approach\end{tabular} &
  \begin{tabular}[c]{@{}l@{}}SAR (search and rescue), \\ Deep Nural Network-SAR (DNN-SAR)\end{tabular} &
  \begin{tabular}[c]{@{}l@{}}Deep dive into DNN-SAR based \\ CBIR workflows and positive \\ outcome on image retrieving\end{tabular} \\ \hline
Noor et al., 2022.\cite{Noor2022-pa} &
  \begin{tabular}[c]{@{}l@{}}Workflow of image \\ retrieving in cloud system\end{tabular} &
  Openstack, Image compression &
  \begin{tabular}[c]{@{}l@{}}Image retrieving technique \\ in cloud.\end{tabular} \\ \hline


Lima et al., 2019.\cite{ref-28} &
  OpenStack Architecture &
  \begin{tabular}[c]{@{}l@{}}Openstack, PRISMA, \\ Horizon, Neutron, Zaqar, \\ Swift, Barbican\end{tabular} &
  A complete view of openstack architecture. \\ \hline
Evans et al., 2018.\cite{ref-30} &
  \begin{tabular}[c]{@{}l@{}}Object Storage searching with indexing\\  and metadata.\end{tabular} &
  \begin{tabular}[c]{@{}l@{}}SQL, Clueso, \\ Amazon Athena, Cassandra\end{tabular} &
  A few more options for searching metadata \\ \hline
Gugnani et al., 2017.\cite{ref-48} &
  \begin{tabular}[c]{@{}l@{}}Find out the common issues of Swift\\  default architecture and designs and\\  new implements.\end{tabular} &
  \begin{tabular}[c]{@{}l@{}}InfiniBand, RoCE \\ such as RDMA\end{tabular} &
  \begin{tabular}[c]{@{}l@{}}Workload increases up to \\ 2x and performance \\ boosted up to 7.3x\end{tabular} \\ \hline
Noor et al., 2017\cite{ref-56} &
  Reliable and secure image processing &
  \begin{tabular}[c]{@{}l@{}}P-Fibonacci transform \\ of Discrete Cosine Coefficients (PFCC), \\ Openstack Swift\end{tabular} &
  \begin{tabular}[c]{@{}l@{}}Performance improvement of image \\ processing and retrieving  with \\ the proposed framework\end{tabular} \\ \hline
Yigal et al., 2016.\cite{ref-33} &
  Monitoring tools for OpenStack &
  \begin{tabular}[c]{@{}l@{}}Openstack, Elasticsearch, \\ Logstash, Kibana\end{tabular} &
  \begin{tabular}[c]{@{}l@{}}A way to monitor and retrieve \\ Openstack's data \\ with middleware tools\end{tabular} \\ \hline
Biswas et al., 2015\cite{ref-26} &
  ACL in OpenStack &
  \begin{tabular}[c]{@{}l@{}}Openstack Swift, Swift ACL, \\ JSON\end{tabular} &
  \begin{tabular}[c]{@{}l@{}}User-level access with ACL feature to \\ retrieve the object.\end{tabular} \\ \hline
Wang et al., 2015\cite{lit-3} &
  Ciphertext based CBIR system &
  \begin{tabular}[c]{@{}l@{}}CKKS (Cheon-Kim-Kim-Song) \\ homomorphic encryption, DNN\end{tabular} &
  \begin{tabular}[c]{@{}l@{}}Using DNN and CKKS, LHS \\ (locally sensitive hash) retrieved \\ image from ciphertext\end{tabular} \\ \hline
Brandt et al., 2003\cite{ref-27} &
  \begin{tabular}[c]{@{}l@{}}How to access large scale of metadata\\  avoiding bottleneck\end{tabular} &
  \begin{tabular}[c]{@{}l@{}}LH (Lazy Hybrid), \\ OBSD, ACL\end{tabular} &
  \begin{tabular}[c]{@{}l@{}}High performance cluster with \\ scalability and flexibility.\end{tabular} \\ \hline
\end{tabular}%
}
\caption{Noteworthy findings from literature review}
\label{table:lit_review_findigs}
\end{table*}

\subsection{Query Searching}
Searching in object storage is now common in cloud systems. Through our studies and findings, we try to find out the drawbacks, and issues of searching, and how they can be solved. A study \cite{intro-1} describes that Swift is a proxy server-based design that has the scale-ability of clusters. 
They propose a change in Swift's architecture that will provide much faster bandwidth with minimal latency while interacting with technology like RoCE, InfiniBand, and Remote Direct Memory Access (RDMA) \cite{ref-48}. 

Imran et al. \cite{intro-2} present some probable problems with metadata-related issues that we may face. In cloud storage, a lot of metadata are created which hampers the performance of the system. They propose an optimized solution for storing massive metadata which has improved load balancing modules and merges storage facility. Xue et al. \cite{ref-49} use HAproxy and UCARP to handle when huge amounts of metadata and it also reduces the buffering and accelerates read and write performances and overall throughput. Metadata is basically stored in a system as small files and with the increasing use of automated technology and remote sensing technology, lots of metadata is produced every day. 

Biswas et al. \cite{ref-26} 
show how Access Control List (ACL) maintains accessibility and data security for all users. With ACL, it can be described who to give access to or not. 
As for the storing policies they make, two types of policies for two types of data. One is LaBAC for user label data and objects label values, another is content level for JSON paths and labels. They find a drawback to their work, stating as it can only work with objects with applications or in JSON. Objects without a JSON file create issues with sending the full content of the files without requesting. 

          

\begin{figure*}
\centering

\subfloat[Overview of the storage architecture \cite{ref-56}]
    {\frame{\includegraphics[width=0.49\linewidth]{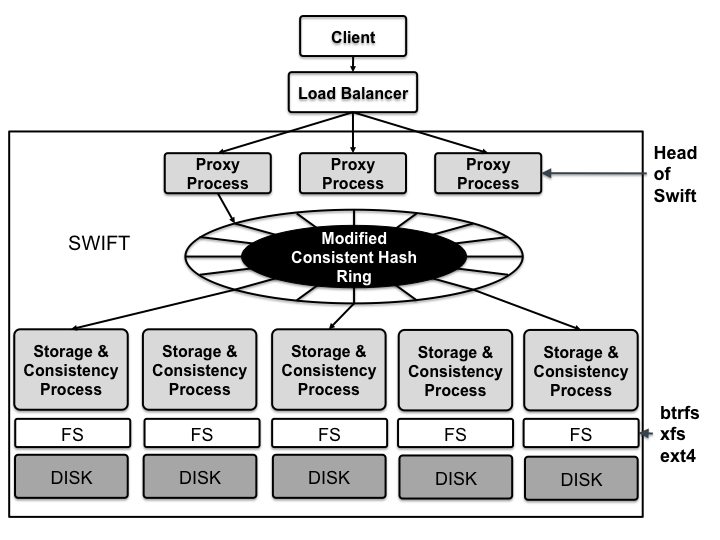}}
     \label{fig:Overview of the storage architecture}}
     \hspace*{.1cm}
\subfloat[Different consistency processes and layers in
proxy and storage nodes of OpenStack Swift]
     {\frame{\includegraphics[width=0.49\linewidth]{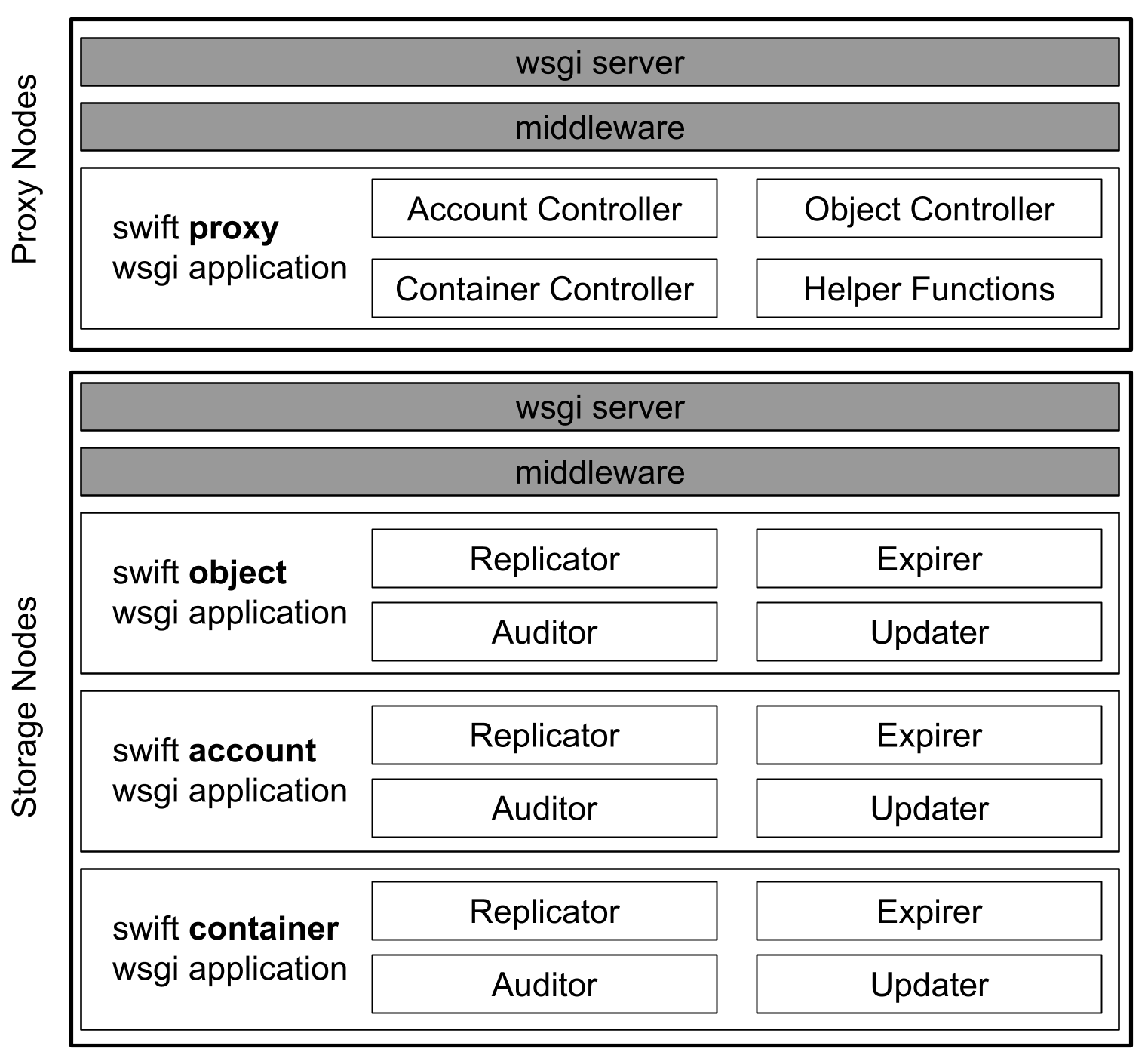}}
     \label{fig:stack}}
     \hspace*{.1cm}
\caption{OpenStack Swift}
\label{fig:architecture_and_processes}
\vspace{-0.8cm}
\end{figure*}

\subsection{Content-based Image Retrieval System}

Ren et al. \cite{Ren_2023} introduce an Approaching-and-Centralizing Network, which can jointly optimize sketch-to-photo synthesis and image retrieval, in which the retrieval module aids the synthesis module in producing large amounts of different photo-like images that gradually approach the photo domain. 

Kakizaki et al. \cite{10030859} introduce a defense approach, to fight against adversarial examples (AXs), that uses deep neural network (DNN) based content-based image retrieval (CBIR). 
Choe et al. \cite{Choe2022-as} propose a CNN based CBIR approach to diagnosing Interstitial Lung Disease with Chest CT. Monowar et al. \cite{Monowar2022-eg} introduce a Deep CNN-based self-supervised image retrieval system. 

Keisham et al. \cite{Keisham2022-ip} present a Deep Search and Rescue (SAR) Algorithm-based CBIR approach. The steps involved in the proposed Deep Neural Network-SAR (DNN-SAR) are pre-processing, multiple feature extraction, feature fusion, clustering, and classification. 

Schall et al. \cite{schall2021gpr1200} come up with a protocol for testing deep learning based models for their general-purpose retrieval qualities. After analyzing the currently existing and commonly used evaluation datasets they conclude with the result that none of the available test sets are suitable for the desired purpose and present the GPR1200 (General Purpose Retrieval) test set. 

Wang et al. \cite{WANG2023103163} propose a secure and efficient ciphertext image retrieval scheme based on image content retrieval (CBIR) and approximate homomorphic encryption (HE). 

Noor et al. \cite{Noor2022-pa} propose a novel approach to retrieve images faster by customizing the attributes in bit pixels of distinct luma and chroma components (Y, Cb, and Cr) of progressive JPEG images. 


\subsection{Keyword Extraction from Document}
Researchers \cite{https://doi.org/10.48550/arxiv.2106.04939} present a multimodal key-phrase extraction approach, namely Phraseformer, using transformer and graph embedding techniques. 
Xiong et al. \cite{Xiong2021-bz} propose Semantic Clustering TextRank (SCTR), a semantic clustering news keyword extraction algorithm based on TextRank that uses BERT to perform k-means clustering to represent semantic clustering. Then, the clustering results are used to construct a TextRank weight transfer probability matrix. Finally,
Iterative calculation of word graphs and extraction of keywords is performed. 

The recent solutions for searching in object storage and their findings are presented in Table \ref{table:lit_review_findigs}. Their drawbacks inspire us to come up with a new solution with the help of object detection and natural language processing that is robust. 
To the best of our knowledge, our proposed methodology
is the first to focus on these aspects.

\section{Background}\label{background}

This section goes over the fundamental architectural framework of Swift, YOLO, BERT, and Elasticsearch.

\subsection{Architectural Overview of Swift}

OpenStack Swift is a highly scalable object storage that is designed keeping in mind the phrase that ``failure is a common occurrence". As a result, Swift is divided into 4 subsections: Proxy, Account, Container, and Object nodes. A proxy server is located in the first layer. Data that goes in and out of the storage has to go through the HTTP file transfer protocol. The requests for data are done by API requests. The task of the proxy server is to capture the requests and work accordingly. The proxy server determines the location of the data or its storage node by the URL. There are Rings, which keep the address of the information like names and entries that are stored on the cluster. It also keeps track of the path of the data. The way Rings keep the mapping work is by introducing zones, devices, partitions, and replicas. Zones can be any storage device, for instance, a hard drive to a full server. After that, there are containers and accounts. The list of containers in a particular account is stored in that account's database. Swift has multiple object nodes which are independent of each other. These object nodes are easily replaceable in the event of any failure. However, Swift has an internal replication system that replicates the stored object into a minimum of three different nodes. So when one node is replaced the objects are not lost\cite{ref-28}. Figure \ref{fig:Overview of the storage architecture} presents the architectural overview of OpenStack Swift and Figure \ref{fig:stack} presents the different consistency processes and layers in proxy and storage nodes of OpenStack Swift.

\begin{figure}[!h]
\centering
\includegraphics[width=0.6\columnwidth]{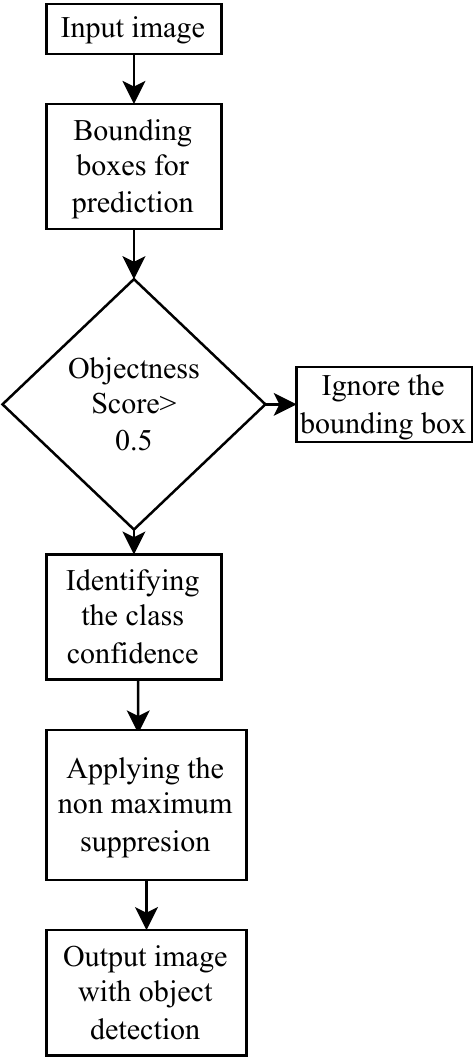}
\caption{Workflow of YOLOv4}
\label{fig:YOLOv4 Workflow}
\vspace{-0.8cm}
\end{figure}

\begin{figure*}[!ht]
\centering
\includegraphics[width=0.7\linewidth]{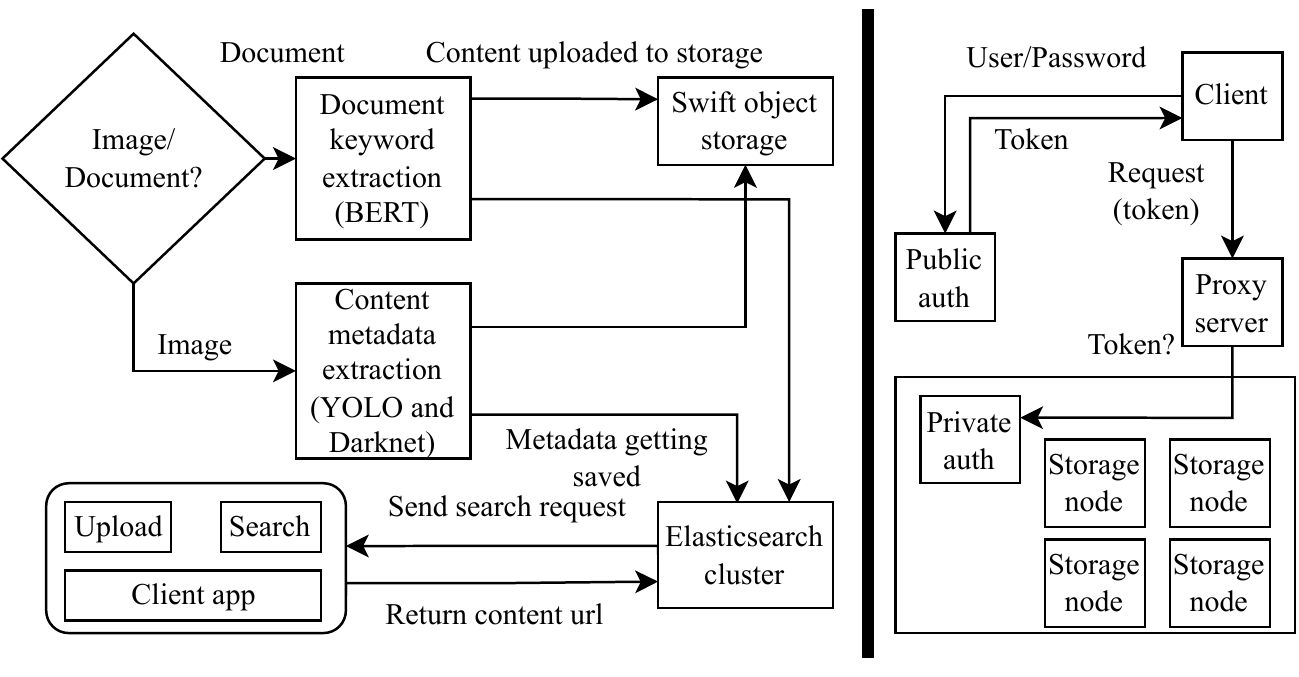}
\caption{Overview of the proposed architecture [on left] and Java library for OpenStack Swift (JOSS) [on right]}
\label{fig:Overview of the proposed system}
\vspace{-0.8cm}
\end{figure*}



\subsection{YOLOv4}
YOLOv4 \cite{yolov4}, \textit{You Only Look Once} Version 4, is a sophisticated technique for single-stage object detection that utilizes regression techniques to gain a good precision score and can be performed concurrently, It is the predecessor of the YOLO line of algorithms. This algorithm is introduced in 2020 by Bochkovskiy et al. \cite{yolov4} and builds upon the capabilities of previous versions such as YOLOv1, YOLOv2, and YOLOv3., achieving the finest potential balance between detection efficiency and precision at this time. Its architecture, which includes the backbone, the neck, and the prediction, is shown in Figure \ref{fig:YOLOv4 Workflow}. 


The authors suggested the following backbones for the YOLOv4 object detector: CSPResNext50, CSPDarknet53, and EfficientNet-53 \cite{yolov4}. As for our purpose, we use CSPDarknet53.
In YOLOv4, the CSPDarkNet53 architecture is introduced. This architecture includes a residual module where the feature layer is re-entered, resulting in increased feature information. This entitles the model to learn the distinction between output and input. 

\subsection{YOLOv8}
YOLOv8 is the successor of all the previous YOLO models presented. It is introduced by Jocher et al. \cite{jocher2023yolo}. YOLOv8 has an architecture similar to one of its ancestors, YOLOv5. 
It is based on PyTorch. And it has a Python backbone rather than Darknet, which is based on C used in YOLOv4. This is convenient for the users to make it customizable and bring improvements through the model. 

On the MS COCO dataset
, YOLOv8m achieve an AP of 53.9\% with a 640-pixel image size (compared to 50.7\% for YOLOv5 on the same input size) at a speed of 280 FPS on an NVIDIA A100 and TensorRT \cite{ACompreh54:online}.



\subsection{BERT}

Bidirectional Encoder Representations from Transformers (BERT) architecture is based on a multi-layer Transformer encoder, developed by Vaswani et al. \cite{Vaswani2017-yz}. Devlin et al. \cite{Devlin2018-sw} present BERT Transformer, which is based on bidirectional self-attention. This bidirectional process eliminates the limitation that self-attention may only integrate context from one side: the left or the right. Unlike previous embedding generation architectures, such as Word2Vec \cite{Mikolov2013-wj}, BERT does not take input vectors that represent words. Instead, it takes segment, token, and position embeddings as input. The token embedding is a WordPiece embedding with 30,000 tokens \cite{Wu2016-kw}.


In this study, we employ the fundamental BERT model, which is available on TensorFlow Hub. It includes 12 transformer blocks, 12 self-attention heads, and 768 hidden size.




\subsection{ElasticSearch Overview}
Elasticsearch is a full-text search library based on the open-source search engine Apache Lucene. It is capable of performing a full-text search. It can conduct a structured search, analytics, or a combination of all three as this is built for real-time, distributed search and data analysis \cite{ref-47}. The highly adaptable query API of Elasticsearch allows for the simultaneous use of filtering, sorting, pagination, and aggregation in a single query \cite{ref-44}. Elasticsearch is capable of easily handling unstructured data, and allowing for the indexing of JSON files without the need for a prior schema. It automatically attempts to identify class mappings and adjusts for new or removed fields. It also offers built-in functionality for clustering, data replication, and instantaneous fail-over, all of which are transparent to the user \cite{ref-44}.

\section{System Design and Implementation}\label{method}

Figure \ref{fig:Overview of the proposed system} (on left)
presents the methodology of our proposed system. First, it checks if the file is an image or a document. Based on the file type, the content is sent to extract the crucial information. For images, the content is sent to extract the metadata. And for documents, the content is inputted to extract the keyword. YOLO (for image) and BERT (for document) process the metadata and send the data to the Elasticsearch cluster. The object detection/keyword extraction is done on the client app. When the metadata is uploaded in the Elasticsearch cluster, the content is uploaded to the Swift server.

\begin{figure}[!h]
\centering
\includegraphics[width=0.75\columnwidth]{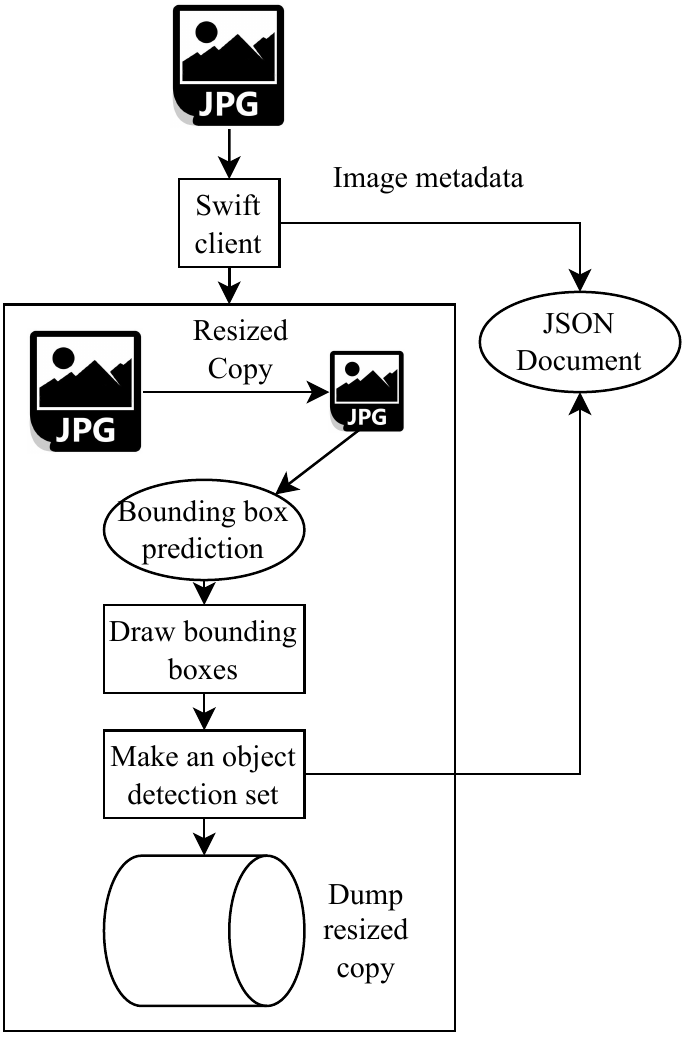}
\caption{Overview of the modified YOLOv4 and YOLOv8 object detector}
\label{fig:Overview of the modified YOLOv4 Object Detector}
\vspace{-0.8cm}
\end{figure}

\subsection{Developing Client-side}
We use Java client for OpenStack Swift (JOSS)\cite{ref-50}, mentioned in the right side of Figure \ref{fig:Overview of the proposed system}, to build the client app. We use Elasticsearch as it offers multi-language support for handling request and response data, language detection libraries, and plugins and integrations to provide additional language-specific functionality. And our JOSS client connects well with Elasticsearch. The location path of the content in our storage server is saved in the Elasticsearch cluster.
We use Elasticsearch as it offers multi-language support for handling request and response data, language detection libraries, and plugins and integrations to provide additional language-specific functionality. And our JOSS client connects well with Elasticsearch.  
When the user searches for content, the client app performs a search in the Elasticsearch cluster and returns content suggestions to the user with the Swift location path. This ensures minimum load on the Swift server and accurate searching based on the metadata. Because of this extraction, the proposed system has a sound knowledge of each content.

OpenStack has a few libraries to interact with the Swift object storage system\cite{intro-1}. We use the Java library for Openstack Swift (JOSS)\cite{ref-50} to develop our client app. It is a desktop-based application where we have our object detection model as well. Object detection is done using the client device's computational power. Then we upload the metadata data to the Elasticsearch cluster and the content to the Swift server. JOSS provides many features to interact with the Swift servers including authentication, object uploading, content location path generation, and so forth \cite{ref-51}.

\begin{figure*}[!h]
\centering
\includegraphics[width=0.6\textwidth]{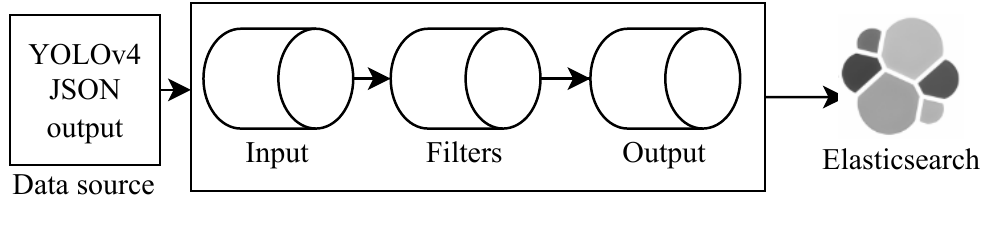}
\caption{Workflow for JSON document in Elastic Cluster}
\label{fig:Workflow for JSON document in Elastic Cluster}
\end{figure*}

\subsection{Developing Keyword Extraction}
BERT is one of the state-of-the-art models to solve problems related to Natural Language Processing (NLP) which uses attention-based mechanisms. In our case, we take a document (docx/pdf) and then take all the text from that document and put it in the BERT model and we the get best five three-word keywords out of the document we input. 

\subsection{Developing Object Detection}
Figure \ref{fig:Overview of the modified YOLOv4 Object Detector} shows how our proposed YOLOv4 algorithm works after it gets an image. YOLOv4 provides us with fast and accurate object detection with the help of bounding boxes and non-maximum suppression. However, in our case, we do not want an edited image with bounding boxes. As a result, we propose a different workflow for the YOLOv4 where after getting an image it will make a copy of that image and perform necessary detections. 
We also followed the same approach for the latest YOLOv8 too And we used a pre-trained YOLOv8m like we used a pre-trained model for YOLOv4 which is already trained with the MSCOCO-17 dataset which has 80 classes of data. 

At that time, the actual image will be sent directly to the storage server. And after that, the detection is done the copied image will be dumped and the object detection set with other metadata related to the image including the object URL path will be written into a JSON document. Lastly, the JSON document will be pushed into the Elasticsearch server.

\subsection{Developing the Storage System}
Figure \ref{fig:Overview of the storage architecture} shows the overall architecture of the storage system. After Detection is done and the object is pushed to Swift, it goes into Swift using the Swift proxy pipeline. In our model, we use a multi-node Swift setup. There are 3 proxy servers and multiple object servers. The load balancer chooses the suitable proxy server for the object. The proxy server sends the object to the Ring, from where the object is sent to the appropriate object server. In our model, we do not change how Swift handles these requests in order to maintain its scalability and compactness.

\subsection{ElasticSearch Cluster}
Figure \ref{fig:Workflow for JSON document in Elastic Cluster} shows the workflow for JSON document in Elastic cluster. We set up an Elasticsearch server in a different Virtual Machine with a Logstash pipeline where the JSON file generated by the object detector gets pushed. Our Logstash pipeline filters out the unnecessary data from the JSON file such as the coordinates of the bounding boxes, and class id. It also formats the JSON file in a way that is easier for Elasticsearch to index properly based on the image file name and the contents of the image which in our case are the detected objects.

\section{Performance Evaluation}\label{results}

\begin{figure}[!h]
\centering
\includegraphics[width=0.65\columnwidth]{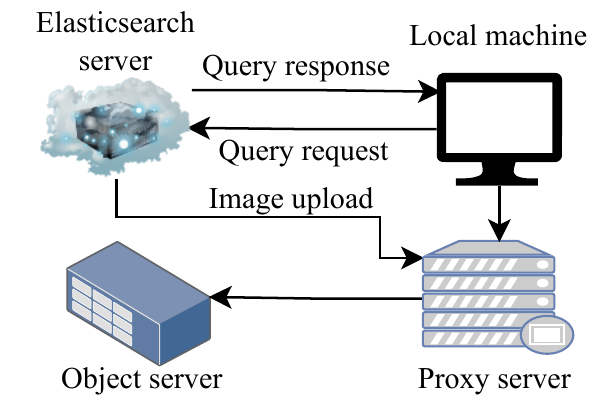}
\caption{Experimental setup overview}
\label{fig:Experimental Setup Overview}
\end{figure}
We measure the performance of our suggested architecture using a real-world implementation. Before this, we first elaborate on proposed experimental settings.

\subsection{Experimental Setup}

We set up multiple virtual machines using the Google Cloud Platform (GCP). One machine work as a proxy server and the other one as the object server node. Account and container servers are included in the object server machines which are shown in Figure \ref{fig:Experimental Setup Overview}. We use our local machine to detect and upload images for this testing. The configuration for our local machine is as follows: Intel(R) Core i5-7300HQ CPU having a 2.50 GHz base clock speed, 8 GB ram, and an Nvidia GTX 1050 Graphics Unit. 

Next, we use another graphics unit Nvidia RTX 2070 where we run our testbed with YOLOv8.

Furthermore, we use another virtual machine as the Elasticsearch server, which is configured to get data dumps from our local machine. For all the Virtual machines, we use the same kind of setup: 4 GB ram, 60 GB storage, and the operating system is Ubuntu 18.04.

\subsection{Dataset}
For our object detection module, we use the Microsoft COCO Dataset
\cite{ref-53}. We take a part of the dataset which consists of 26,000 images and 80 classes in total. After, we divide our dataset into three segments for our testing purposes having 1000, 5000, and 20000 Images each. 
 We use pre-trained YOLOv4 and
YOLOv8m models 
to test these images. 




For our keyword extraction, we use the SemEval2017 Dataset \cite{Augenstein2017-lq} which consists of paragraphs selected from 500 ScienceDirect journal papers from Computer Science, Material Sciences, and Physics domains.

\subsection{Experimental Results}

In this section, we report the results after using the datasets in our system, starting with the image dataset, and afterward the document dataset.

\begin{figure*}
\centering

\subfloat[Aeroplane]
    {\frame{\includegraphics[width=0.24\linewidth]{figures/aeroplane.jpg}}
     \label{result_img_1}}
     \hspace*{.1cm}
\subfloat[Aeroplane (Detected)]
     {\frame{\includegraphics[width=0.24\linewidth]{figures/aeroplane_detected.jpg}}
     \label{result_img_2}}
     \hspace*{.1cm}
\subfloat[Man in Horse]
    {\frame{\includegraphics[width=0.24\linewidth]{figures/man_in_horse.jpg}}
     \label{result_img_3}}
\subfloat[Man in Horse (Detected)]
    {\frame{\includegraphics[width=0.24\linewidth]{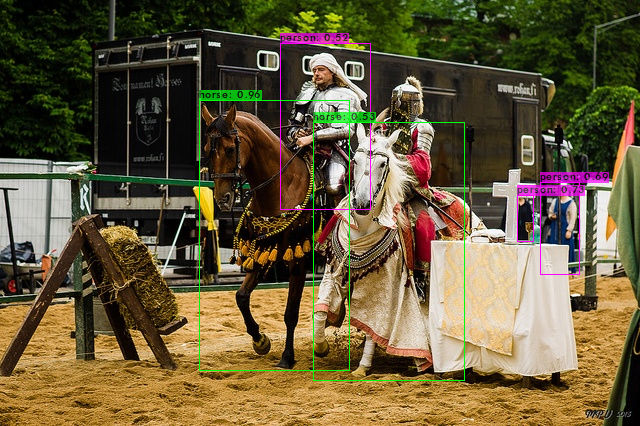}}
     \label{result_img_4}}
\caption{Object Detection (Single and Multiple)}
\label{fig:object detection}
\vspace{-0.8cm}
\end{figure*}

\textit{Image Dataset Test.}
From Table \ref{dataset_testing_metrics}, we can see different matrices from our testing sets. The precision level is important for our model because it indicates how well our system is able to give the user the appropriate image they want. In our case, we get an mAP (Mean Average Precision) of 0.71 for 5k images and 0.73 for 20 thousand images with a total precision of 68.5\%. In Figure \ref{fig:object detection}, we can see both single and multiple objects are being detected by the model. 


\begin{table}[]
\centering

\begin{tabular}{|l|l|l|}
\hline
 & 5000 Images & 20000 Images \\ \hline
Detection time (sec) & 681 & 2907 \\ \hline
mAP & 0.71 & 0.73 \\ \hline
Average IoU & 0.55 & 0.56 \\ \hline
F1 score & 0.68 & 0.69 \\ \hline
Recall & 0.68 & 0.69 \\ \hline
Precision & 0.68 & 0.69 \\ \hline
\end{tabular}
\caption{Dataset testing metrics}
\label{dataset_testing_metrics}
\end{table}

\begin{figure*}
\centering
\subfloat[Detection time (seconds)]
    {\includegraphics[width=0.32\linewidth]{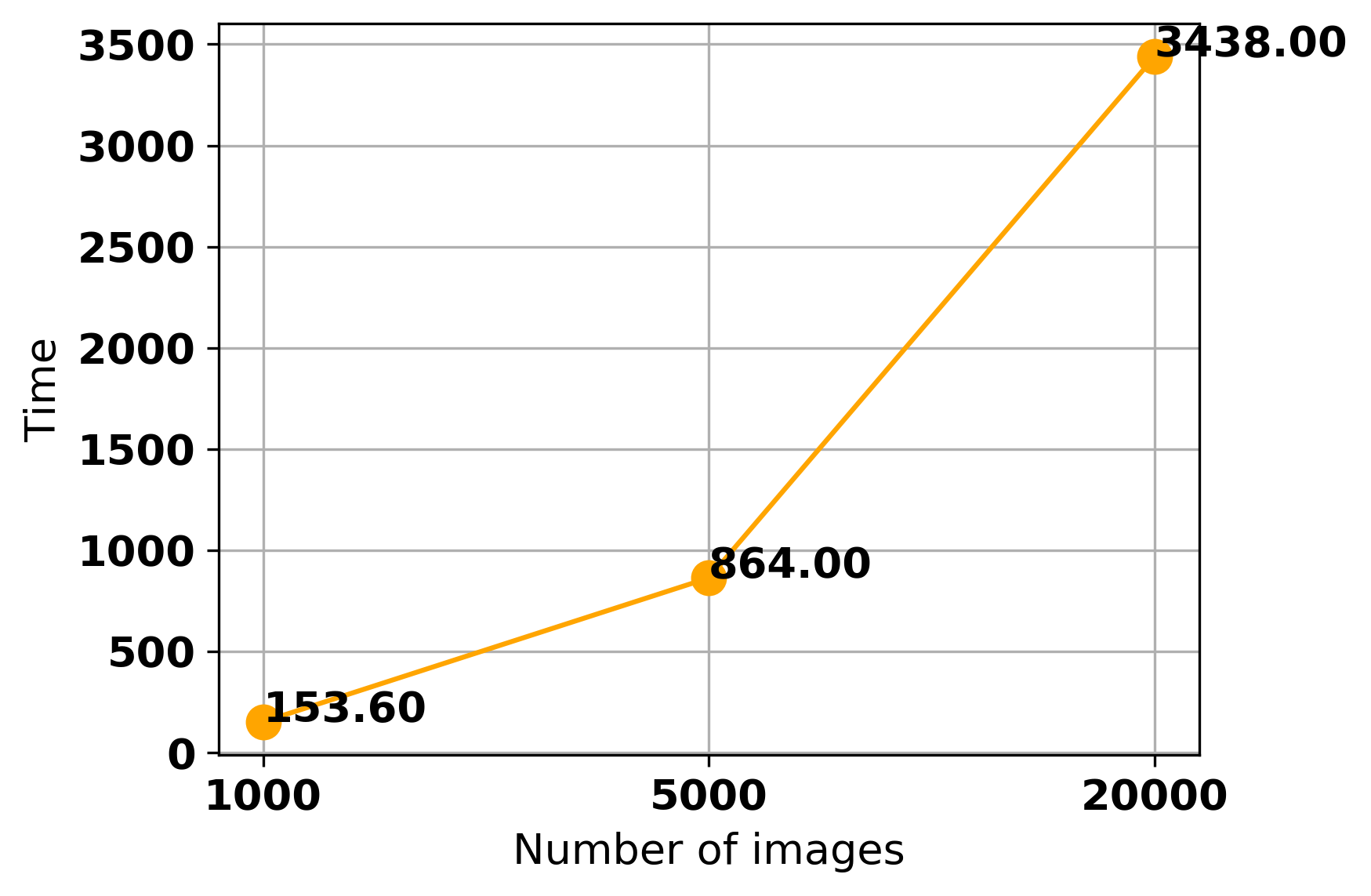}
     \label{fig:Detection Time graph v4}}
\hspace{0.01in}
\subfloat[Upload time (seconds)]
    {\includegraphics[width=0.32\linewidth]{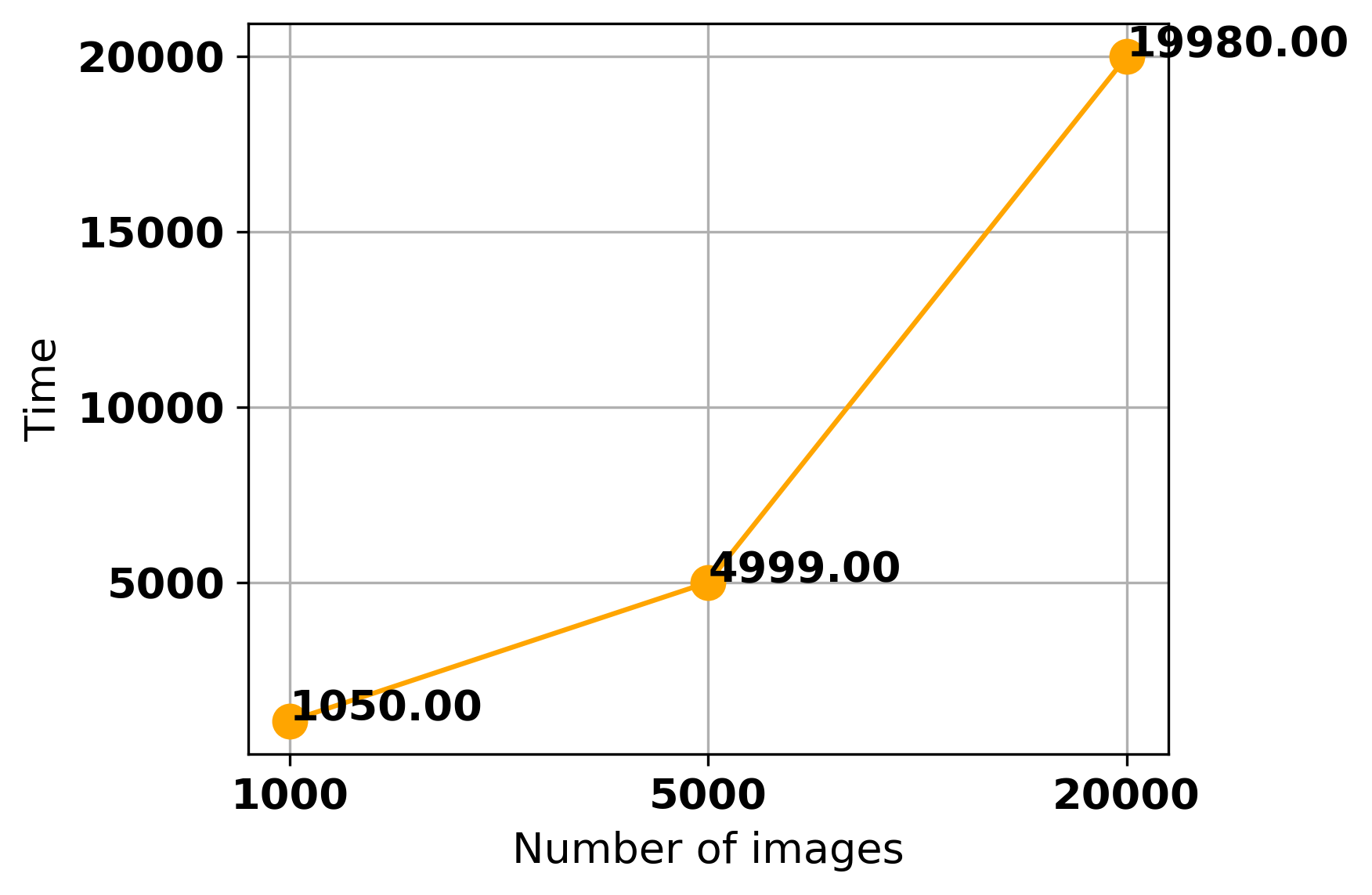}
     \label{fig:Upload Time Graph v4}}
\hspace{0.01in}
\subfloat[Upload \& detection time (seconds)]
    {\includegraphics[width=0.32\linewidth]{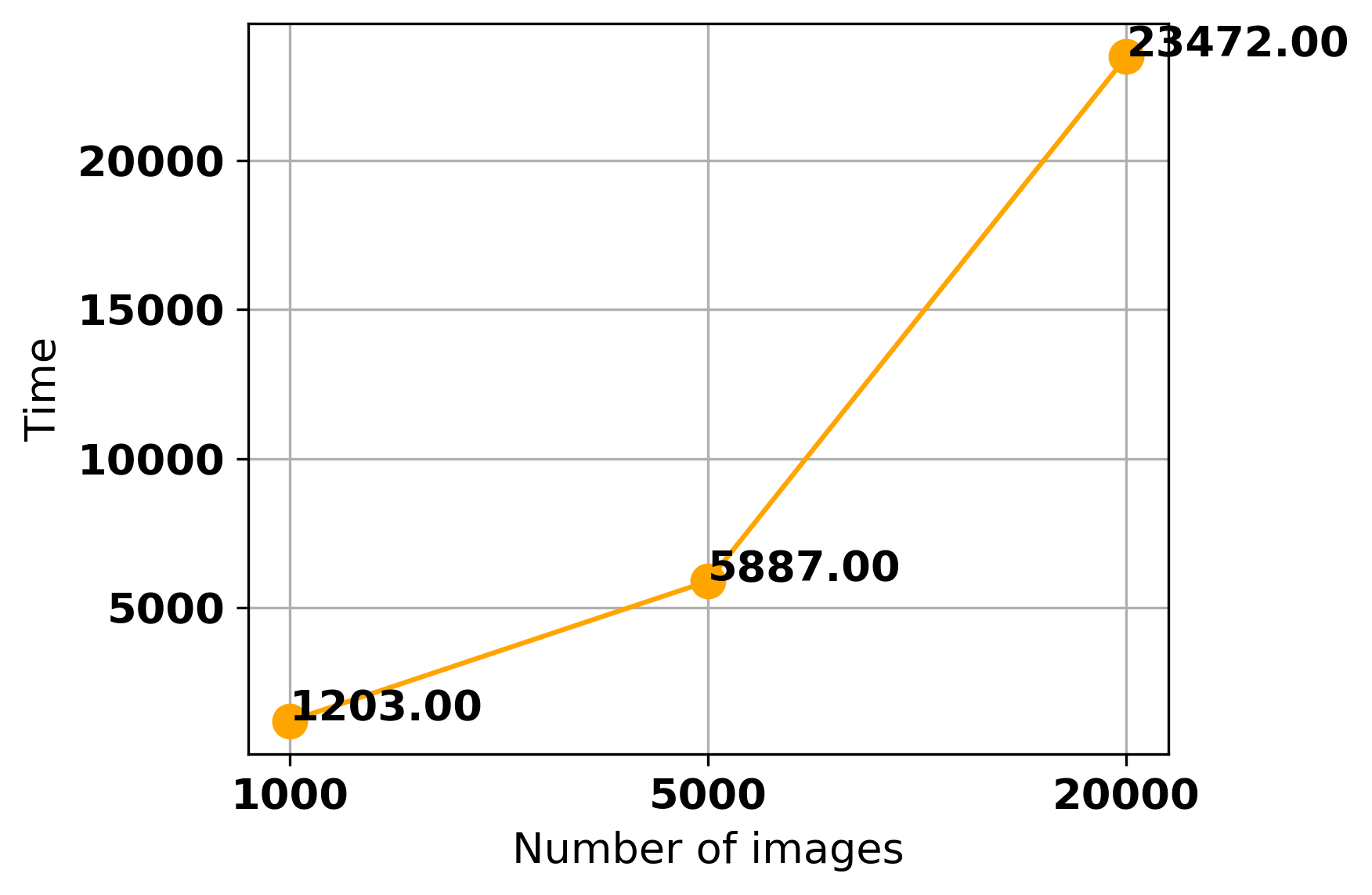}
     \label{fig:Upload & Detection Time Graph YOLOv4}}\\
\subfloat[Detection time (seconds)]
    {\includegraphics[width=0.32\linewidth]{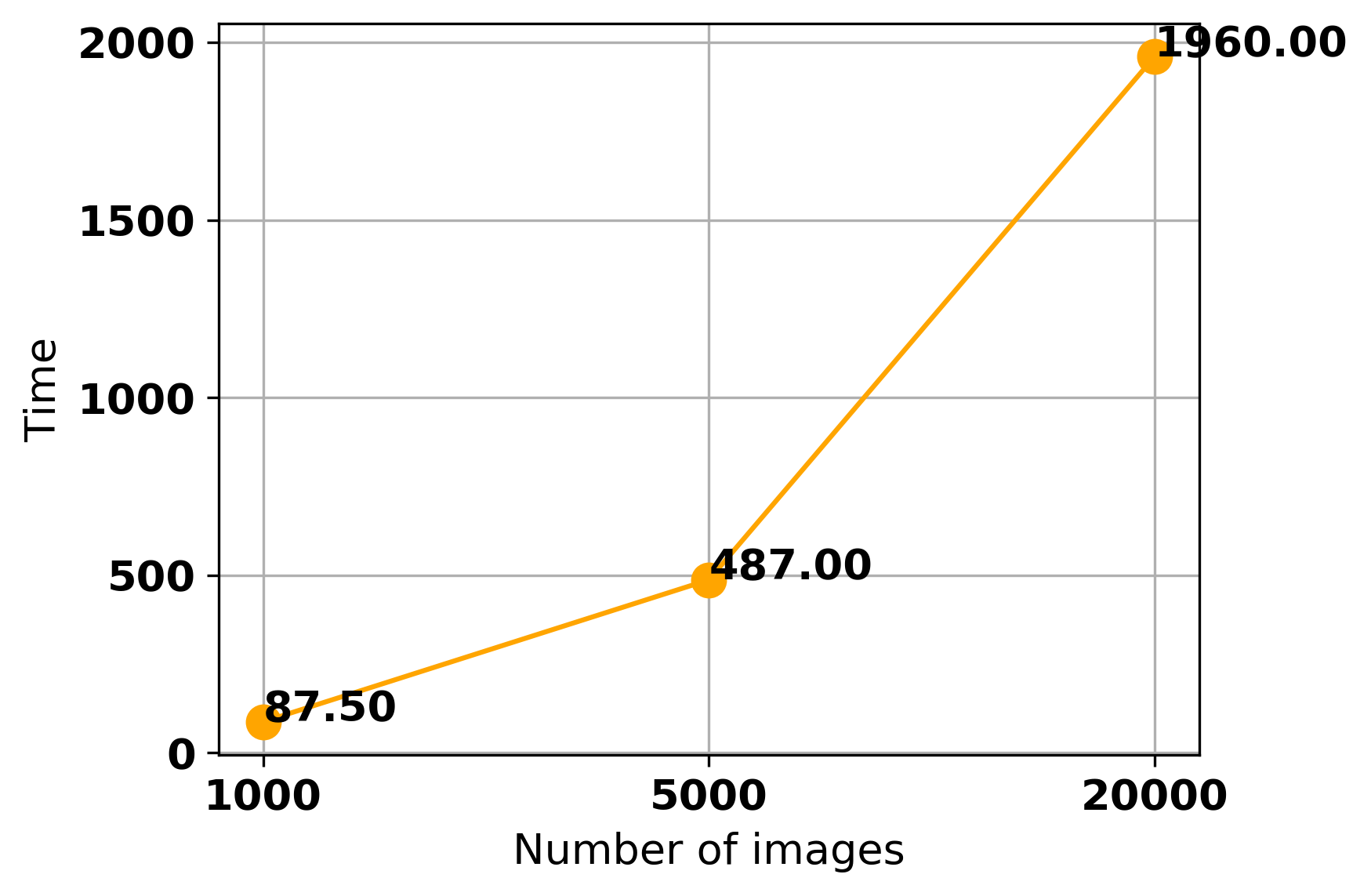}
    \label{fig:Detection Time graph v8}}
\hspace{0.01in}
\subfloat[Upload time (seconds)]
{\includegraphics[width=0.32\linewidth]{figures/upload_time_new.png}
     \label{fig:Upload Time Graph v8}}
\hspace{0.01in}
\subfloat[Upload \& detection time (seconds)]
    {\includegraphics[width=0.32\linewidth]{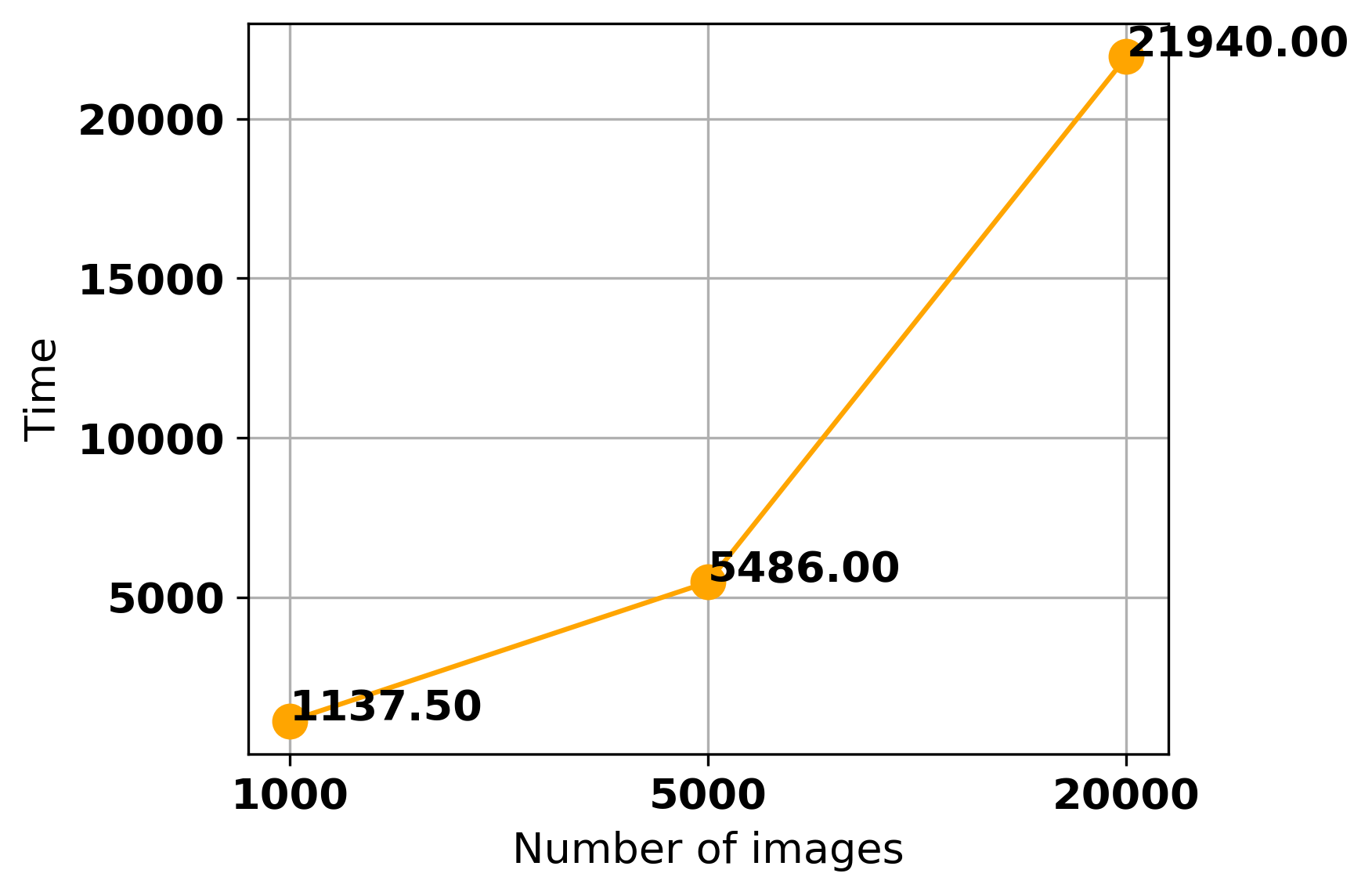}
     \label{fig:Upload & Detection Time Graph YOLOv8}}

\caption{Time graph for YOLOv4 (a, b, c) and YOLOv8 (d, e, f)}
\label{yolo_time_graph}
\vspace{-0.5cm}
\end{figure*}

\begin{table}[!h]
\centering
\begin{tabular}{|l|l|l|l|}
\hline
\begin{tabular}[c]{@{}l@{}}Image \\ count\end{tabular} &
  \begin{tabular}[c]{@{}l@{}}Average\\ detection\\ time\\ (YOLOv4)\end{tabular} &
  \begin{tabular}[c]{@{}l@{}}Average\\ upload\\ time\end{tabular} &
  \begin{tabular}[c]{@{}l@{}}Total\\ upload \&\\ detection\\ time\\ (YOLOv4)\end{tabular} \\ \hline
1000  & 153.6 & 1050  & 1204  \\ \hline
5000  & 864   & 4999  & 5887  \\ \hline
20000 & 3438  & 19980 & 23472 \\ \hline
\end{tabular}
\caption{Average time to do different tasks for different data sizes with YOLOv4 (Time = second)}
\label{tab:Average Upload & Detection time for different data sizes with YOLOv4}
\end{table}

\begin{table}[!h]
\centering
\begin{tabular}{|l|l|l|l|}
\hline
\begin{tabular}[c]{@{}l@{}}Image \\ count\end{tabular} &
  \begin{tabular}[c]{@{}l@{}}Average\\ detection\\ time\\ (YOLOv8)\end{tabular} &
  \begin{tabular}[c]{@{}l@{}}Average\\ upload\\ time\end{tabular} &
  \begin{tabular}[c]{@{}l@{}}Total\\ upload \&\\ detection\\ time\\ (YOLOv8)\end{tabular} \\ \hline
1000  & 87.5 & 1050  & 1138  \\ \hline
5000  & 443  & 4999  & 5442  \\ \hline
20000 & 1790 & 19980 & 21770 \\ \hline
\end{tabular}
\caption{Average time to do different tasks for different data sizes with YOLOv8 (Time = second)}
\label{tab:Average Upload & Detection time for different data sizes with YOLOv8}
\vspace{-0.8cm}
\end{table}

\textit{Detection time test.}
Figure \ref{fig:Detection Time graph v4} and \ref{fig:Detection Time graph v8} represents the different times it takes to detect 1,000, 5,000, and 20,000 images. We can see a very low upward-sloping curve which tells us the detection time is very little compared to the number of images. We are able to achieve this because of the YOLO algorithm which as its name suggests, \textit{ONLY LOOKS AT AN IMAGE ONCE}. However, removing the bounding box drawing method only increases the speed very slightly which can be overlooked. 


\textit{Upload time test.}
In the uploading part, we limit our upload speed to 2mbps (megabit) and calculate the upload time of the images. In Figure \ref{fig:Upload Time Graph v4} and \ref{fig:Upload Time Graph v8}, we get a relatively higher curve because of the low upload speed compared to the file sizes of the images.


\textit{Total time for proposed model.}
After we calculate the Detection \& Upload time separately, we go ahead and initialize our system to find out the combined time it needs for uploading and detecting the different image sets. From Figure \ref{fig:Upload & Detection Time Graph YOLOv4} and \ref{fig:Upload & Detection Time Graph YOLOv8}, we can see the curve going higher ever so slightly. 

\begin{figure}[!h]
\centering
\includegraphics[width=.8\columnwidth]{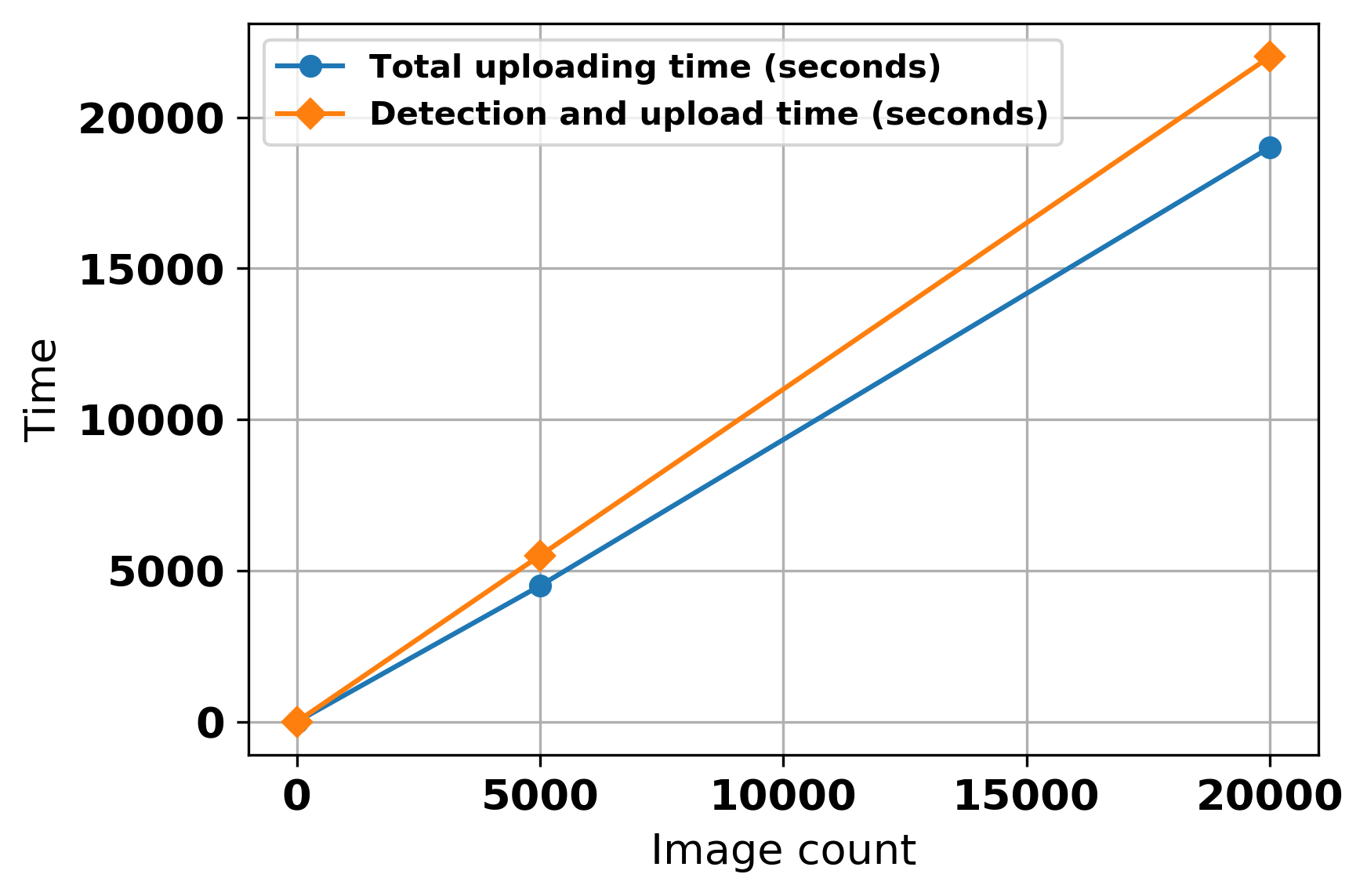}
\caption{Comparison graph}
\label{fig:Comparison Graph}
\vspace{-0.8cm}
\end{figure}

\textit{Uploading and detection time comparison.}
From Figure \ref{fig:Comparison Graph}, we can see the comparison between the Uploading time and the Detection-Upload combined time, indicating that, the additional requirements of object detection and our system necessitates take a little more time to deliver the image to Swift. But the difference is relatively insignificant compared to the work done behind the scenes.
One important thing to note is that we use a comparatively older graphical processing unit model, Nvidia GTX 1050 which is without CUDNN functionality, and as a result, our GPU usage was up to 15-20\% at max. So using a newer version of GPU or even an older version with CUDNN enabled will exponentially decrease the detection time which in turn will decrease the overall upload \& detection time of the system bringing the two curves in Figure \ref{fig:Comparison Graph} much closer to each other. 
Table \ref{tab:Average Upload & Detection time for different data sizes with YOLOv4} and Table \ref{tab:Average Upload & Detection time for different data sizes with YOLOv8} shows the average time our models took to detect \& upload different data segments.

\textit{Result evaluation for document.}

\begin{table}[]
\centering
\begin{tabular}{|l|l|l|l|}
\hline
Document &
  \begin{tabular}[c]{@{}l@{}}Extracted \\ keyword 1\end{tabular} &
  \begin{tabular}[c]{@{}l@{}}Extracted \\ keyword 2\end{tabular} &
  \begin{tabular}[c]{@{}l@{}}Extracted \\ keyword 3\end{tabular} \\ \hline
\begin{tabular}[c]{@{}l@{}}Document A\end{tabular} &
  \begin{tabular}[c]{@{}l@{}}neural \\ networks \\ deep\end{tabular} &
  \begin{tabular}[c]{@{}l@{}}including\\ computer \\ vision\end{tabular} &
  \begin{tabular}[c]{@{}l@{}}deep \\ neural \\ networks\end{tabular} \\ \hline
\begin{tabular}[c]{@{}l@{}}Document B\end{tabular} &
  \begin{tabular}[c]{@{}l@{}}numerical \\ insight \\ recent\end{tabular} &
  \begin{tabular}[c]{@{}l@{}}thermodynamic \\ limit \\ recent\end{tabular} &
  \begin{tabular}[c]{@{}l@{}}theories \\ complex \\ boltzmann\end{tabular} \\ \hline
\end{tabular}
\caption{Extracted keywords from BERT}
\label{tab:bert-document-result}
\vspace{-0.8cm}
\end{table}

Table \ref{tab:bert-document-result} presents the tentative document and the extracted keywords.
We find the documents which are most similar to the document and these are those keywords that can best represent our document. We use cosine similarity between vectors and the document. We select only top three keywords from most similar candidates (documents) to the input document which has three words on each.

\subsection{Search Analysis}

Here, we discuss the functionalities of searching in our system. 

\textit{Completion Suggester.} 
The completion suggester provides auto-complete/search-as-you-type functionality. This navigational feature helps users to get relevant results as they type, improving search precision. 
We achieve completion suggestions with the help of Elasticsearch API, which we integrate with our Client API. 


\textit{Search based on Image Content and Metadata.}
Elasticsearch API uses the Elasticsearch cluster as an endpoint base where all the documents are stored and indexed. Leveraging the API, we successfully implement search based on both image content and metadata.
After we perform a search using 80 keywords (as the dataset has 80 classes), we calculate the average query time and average request time given in the table \ref{tab:Average query time and request time for different data sizes}. We observe that although the average request time fluctuates a little bit, the average query time remains very low and almost constant.


\begin{table}[]
\centering
\begin{tabular}{|l|c|c|c|}
\hline
Documents & \begin{tabular}[c]{@{}c@{}}Total \\ Keywords\end{tabular} & \begin{tabular}[c]{@{}c@{}}Average \\ Query \\ Time(ms)\end{tabular} & \begin{tabular}[c]{@{}c@{}}Average \\ Request\\  Time(ms)\end{tabular} \\ \hline
1000 & 80 & 54 & 440 \\ \hline
5000 & 80 & 57 & 570 \\ \hline
20000 & 80 & 58 & 680 \\ \hline
\end{tabular}%

\caption{Average query time and request time for different data sizes}
\label{tab:Average query time and request time for different data sizes}
\vspace{-0.8cm}
\end{table}

\begin{table*}[h!]
\centering
\begin{tabular}{|l|l|l|l|l|l|}
 \cline{1-5}
{Features} & DevStack & {Traditional Swift} & SwiftStack & \textbf{Proposed} \\   \cline{1-5}
 Deployment & Developer & Developer & Commercial & Developer\\   \cline{1-5}
 \begin{tabular}[c]{@{}l@{}} {Supported Operating Systems} \end{tabular} & \begin{tabular}[c]{@{}l@{}}{Ubuntu, CentOS} \end{tabular} & \begin{tabular}[c]{@{}l@{}}{Ubuntu, Fedora, CentOS} \end{tabular} & \begin{tabular}[c]{@{}l@{}}{Windows, Ubuntu, CentOS}\end{tabular} & \begin{tabular}[c]{@{}l@{}}{Windows, Ubuntu}\\ \end{tabular} \\  \cline{1-5}
 \begin{tabular}[c]{@{}l@{}}{Multi-Node Supports} \end{tabular} & No & Yes & Yes & Yes \\    \cline{1-5}
 OpenSource & Yes & Yes & No & Yes \\   \cline{1-5}
 \begin{tabular}[c]{@{}l@{}}{Stability of Deployed Setup} \end{tabular} & No & \begin{tabular}[c]{@{}l@{}}{Stable than Devstack} \end{tabular} & Yes & Yes \\   \cline{1-5}
 Difficulty to use & Low & High & Low & Low \\   \cline{1-5}
 Setup Difficulty & Low & High & Average & Average \\  
 \cline{1-5}
\end{tabular}
\caption{Difficulty and availability of various implementations of Swift}
\label{difficulty_swift}
\end{table*}

\begin{table*}[h!]
\centering
\resizebox{\textwidth}{!}{%
\begin{tabular}{|l|l|l|l|l|l|l|}
\hline
Author & Dataset & \begin{tabular}[c]{@{}l@{}}Images \& \\ number of classes\end{tabular} & Techniques & Application & Query technique & Precision \\ \hline
Ashraf et al, 2020. \cite{refcp-56} & Corel 1000 & \begin{tabular}[c]{@{}l@{}}100 images \\ with 10 classes\end{tabular} & \begin{tabular}[c]{@{}l@{}}Video, Image data for content-based \\ imagery collection via several methods \end{tabular} & CBIR & Image & 0.875 \\ \hline
Ahmed et  al, 2019. \cite{refcp-58} & Corel 1000 & \begin{tabular}[c]{@{}l@{}}1000 images \\ with 10 classes\end{tabular} & Image features information fusion & CBIR & Image & 0.90 \\ \hline
Nazir et al, 2018. \cite{refcp-63} & Corel 1- K & \begin{tabular}[c]{@{}l@{}}1000 images \\ with 10 classes\end{tabular} & \begin{tabular}[c]{@{}l@{}}HSV analysis of colors, wavelength \\ determination  as well as the edge \\ histogram characterization\end{tabular} & CBIR & Image Text & 0.735 \\ \hline
Mistry et al, 2018. \cite{refcp-57} & Wang & \begin{tabular}[c]{@{}l@{}}1000 images \\ with 10 classes\end{tabular} & \begin{tabular}[c]{@{}l@{}}Hybrid features and various \\ distance metric\end{tabular} & CBIR & Image & 0.875 \\ \hline
Liu et  al, 2017. \cite{refcp-59} & Brodatz & \begin{tabular}[c]{@{}l@{}}1856 and 600 \\ texture images, \\ consisting of 640 \\ and 864 texture \\ images\end{tabular} & \begin{tabular}[c]{@{}l@{}}Fusion of color histogram and \\ LBP-based features\end{tabular} & TBIR & Image & 0.841 \\ \hline
Our Proposed Approach & MS COCO & \begin{tabular}[c]{@{}l@{}}26000 images \\ with 80 classes\end{tabular} & \begin{tabular}[c]{@{}l@{}}Object Detection using Darknet-53 \\ and YOLOv4 \end{tabular} & CBIR & Text & 0.74 \\ \hline
\end{tabular}%
}
\caption{Comparison between various CBIR engines (Here, TBIR = Texture Based Image Retrieval)}
\label{tab:Comparison between various CBIR engines}
\end{table*}

\section{Discussion \& Comparative Analysis}\label{discussion}



The model which we propose is a novel approach to searching for images in OpenStack Swift. As a result, direct comparison with other Swift models is not possible.
And again, it adopts a middleware model to improve searching efficiency and make the entire storage system more object-aware. The middleware increases the speed and effectiveness of searching within the object storage, making it simpler to find and retrieve particular objects. It also increases the storage system's object awareness, giving it more functionality and flexibility for managing and accessing objects. Overall, the proposed solution offers OpenStack Swift users a beneficial improvement, increasing the effectiveness and functionality of their object storage system.
However, we divide our comparison part into two sub-sections. We compare how different swift models and content-based image search models do searching compared to our model based on some parameters.

\textit{Different Swift Models.} 
After using each of the platforms, we set up some parameters for effective image search in swift storage and compare them with our proposed model. From Table \ref{tab:Comparison-Swift}, we can see the comparison of the various implementation of Swift using different techniques for searching. 



\begin{table}[]
{%
\caption{Comparison of various implementations of Swift (Here, \checkmark=Yes and {X}=No)}
\label{tab:Comparison-Swift}
\begin{tabular}{|l|c|c|c|c|}
\hline
Methods & DevStack & \begin{tabular}[c]{@{}c@{}}Traditional \\ Swift\end{tabular} & SwiftStack & Proposed \\ \hline
Metadata & \checkmark & \checkmark & \checkmark & \checkmark \\ \hline
\begin{tabular}[c]{@{}l@{}}Image \\ Content \\ Extraction\end{tabular} & {X} & {X} & {X} & \checkmark \\ \hline
Elasticsearch & \checkmark & {X} & \checkmark & \checkmark \\ \hline
\begin{tabular}[c]{@{}l@{}}Search \\ Optimization\end{tabular} & {X} & {X} & \checkmark & \checkmark \\ \hline
\begin{tabular}[c]{@{}l@{}}Search based \\ on Contents\end{tabular} & {X} & {X} & {X} & \checkmark \\ \hline
\end{tabular}%
}
\vspace{-0.8cm}
\end{table}

In Table \ref{difficulty_swift}, we compare the user availability level of the different implementations based on the model’s availability and scalability to work in different environments.

\textit{Different CBIR Engines.}
There are multiple CBIR engines that extract different features from images to conduct a search. In Table \ref{tab:Comparison between various CBIR engines}, we compare our proposed model with different models from other related works based on what features get extracted and how the search is conducted. Moreover, we compare the precision of these various models. 

When we upload any picture to the server, after inputting the image to YOLOv4 or YOLOv8, the image does not lose it. It shows no change in the image quality (SSIM 100\% and VQMT 100\%) after the images are passed through the detection algorithm.


\section{Conclusion and Future Work}\label{conclusion}

In our work, we integrate machine learning features and OpenStack Swift to come up with a better solution to the problem of efficient searching. With the help of Elasticsearch, we are able to complete the entire design. Although our main objective is to find a solution to the searching method of Swift, we also come up with a secondary objective to make a user-centered content-based image searching \cite{refcp-63} system using a text-based database where a user can manipulate the YOLOv4 and YOLOv8 algorithm based on their preferences, which would neither hinder the performance of the Swift storage nor the Elasticsearch cluster as they are independent of each other. As YOLOv4 and YOLOv8 can do object detection for both images and live video feeds, it adds a variety of choices for different kinds of users.

Different search techniques based on content level metadata are not thoroughly focused on in OpenStack Swift literature. Hence in our paper, we externally integrate an object detection framework and an Elasticsearch cluster to our overall swift storage and perform multiple tests to find out the viability and responsiveness of our model. Although we get results with very little delay, there is still room for improvement. 
As a result, we select three future goals in order to make our system more robust and easy to use. At first, we plan to integrate the whole system more compactly using a desktop-based application. Then, we aim to add an authentication token system to our Elasticsearch server to keep the documents safe from unauthorized access. Afterward, our target is to use our system to store live video feeds in order to find out the viability of our system as a state-of-the-art video surveillance application.






\bibliographystyle{IEEEtran}
\bibliography{reference.bib}


%

\vspace{-1.2cm}
\begin{IEEEbiographynophoto}{Jannatun Noor} has been working as a Senior Lecturer at the Department of CSE, School of Data and Sciences, Brac University. She is the head of Computing for Sustainability and Social Good (C2SG) Research Group at Brac University. Besides, she is working as a Graduate Research Assistant under the supervision of Prof Dr. A. B. M. Alim Al Islam in the Department of CSE at Bangladesh University of Engineering and Technology (BUET). Prior to BRACU, she worked as a Team Lead of the IPV-Cloud team in IPvision Canada Inc. She received her B.Sc. and M.Sc. degrees in Computer Science and Engineering from BUET. Her research work covers Cloud Computing, Wireless Networking, Data Mining, Big Data Analysis, Internet of Things.
\end{IEEEbiographynophoto}

\vspace{-1.2cm}
\begin{IEEEbiography}[{\includegraphics[width=1in,height=1.25in,clip,keepaspectratio]{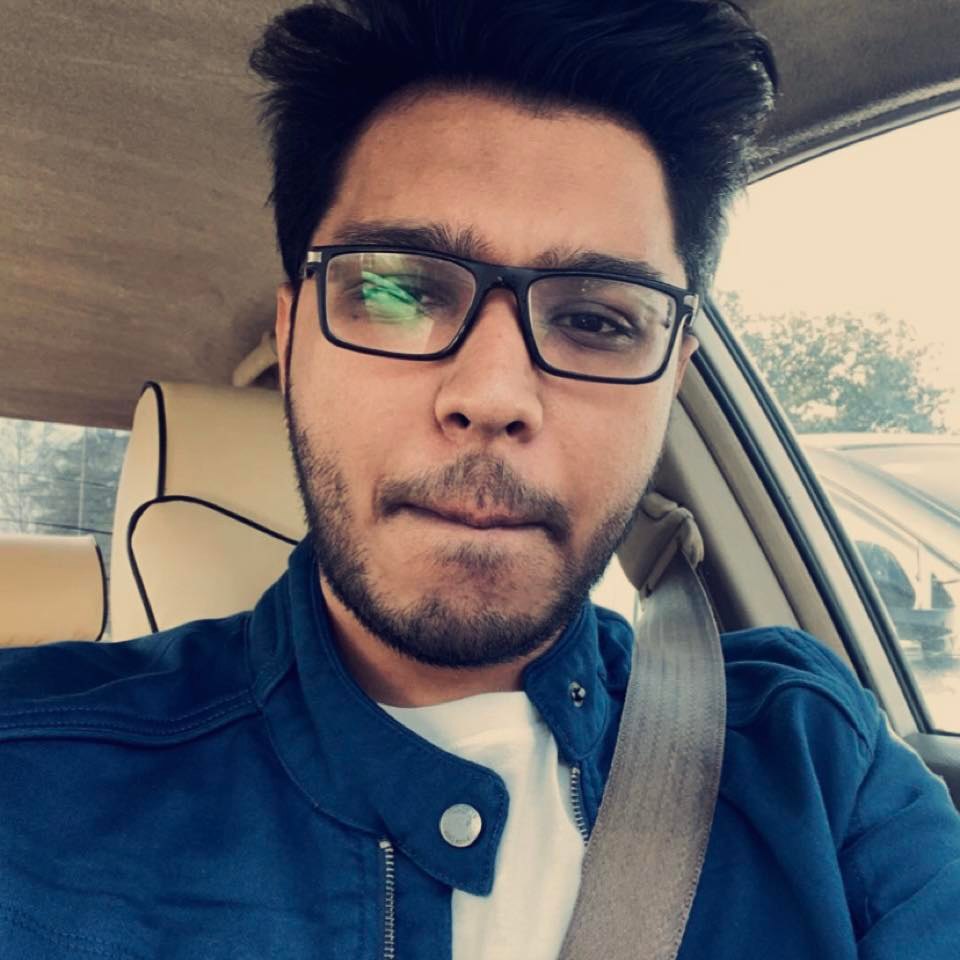}}]
{Rizwanul Haque Ratul} is currently working as a Software Engineer at Optimizely. He completed his Bachelor in Computer Science from Brac University. His research interests include Cloud Computing and Networking.
\end{IEEEbiography}
\vspace{-1.2cm}
\begin{IEEEbiography}[{\includegraphics[width=1in,height=1.25in,clip,keepaspectratio]{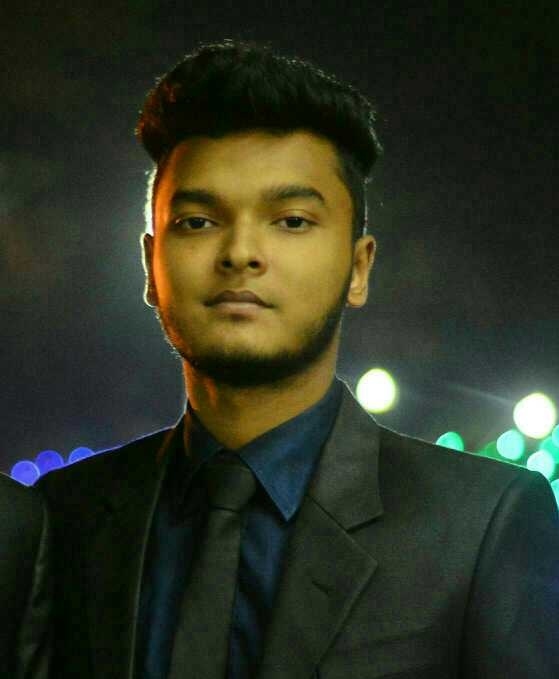}}]{Mir Rownak Ali Uday} has recently completed his Bachelors in Computer Science and Engineering from Brac University, Dhaka, Bangladesh. His research interests include Machine Learning, Deep Learning, Computer Vision, and Cloud Computing.
\end{IEEEbiography}
\vspace{-1.2cm}
\begin{IEEEbiography}[{\includegraphics[width=1in,height=1.25in,clip,keepaspectratio]{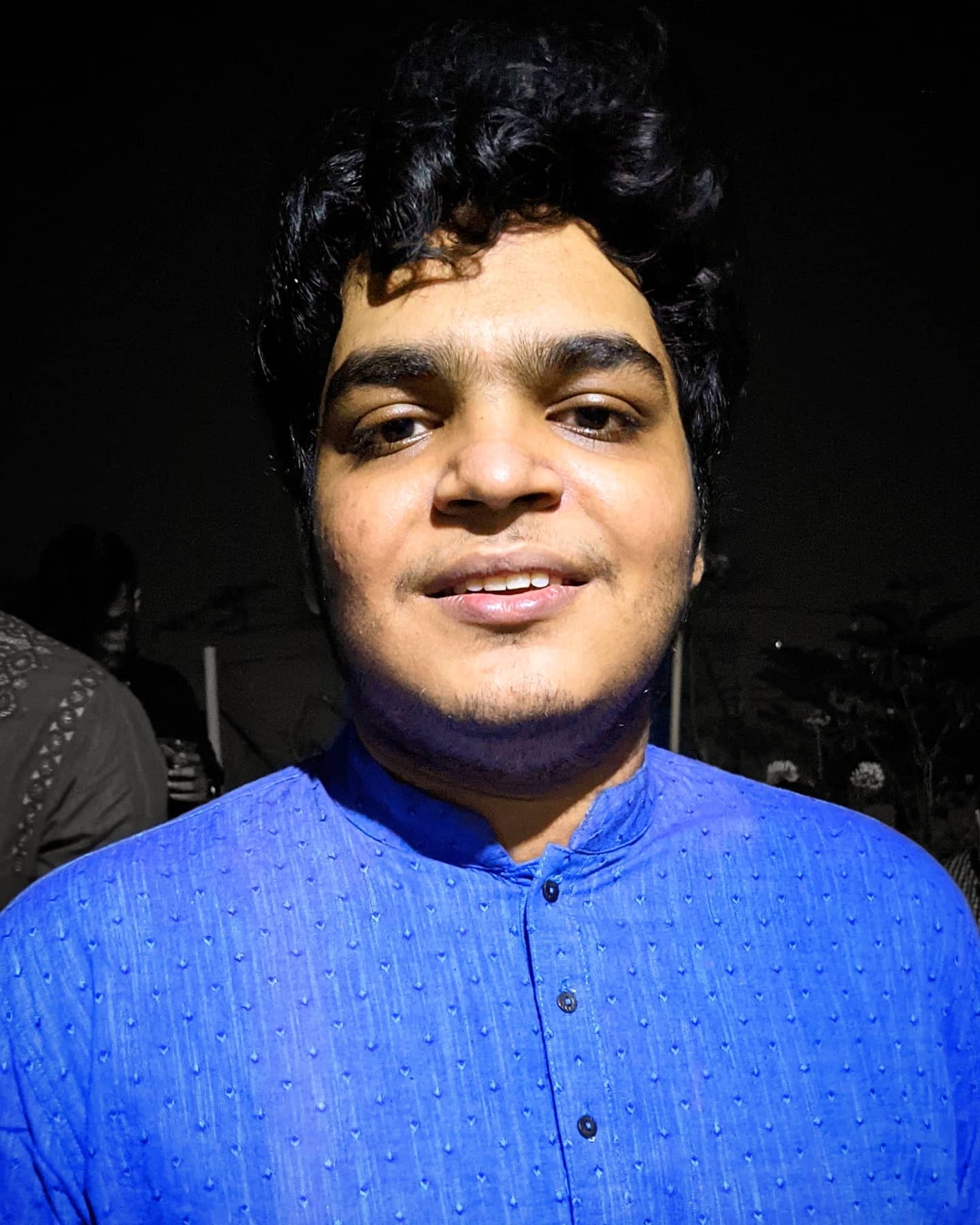}}]{Joyanta Jyoti Mondal} is an incoming graduate student at The University of Alabama at Birmingham. He is also currently working at Computing for Sustainability and Social Good (C2SG) Research Group, School of Data and Sciences, BRAC University. 
His research interests cover Deep Learning, Computer Vision, Cloud Computing, and Modeling \& Simulation. 
\end{IEEEbiography}
\vspace{-1.2cm}

\begin{IEEEbiography}[{\includegraphics[width=1in,height=1in,clip,keepaspectratio]{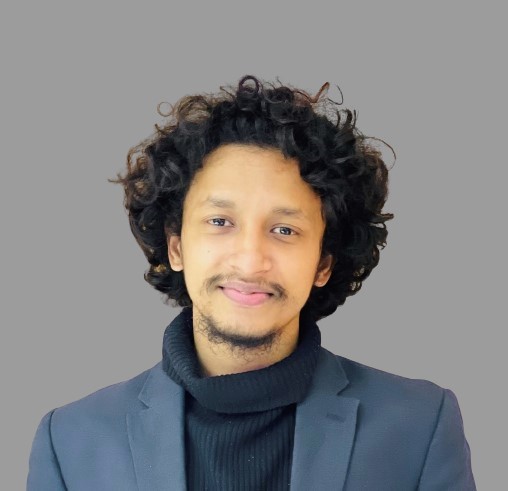}}]{Md. Sadiqul Islam Sakif} is currently working as a Machine Learning Engineer at mPower-Social also pursuing his bachelor's degree from the School of Data and Sciences, Brac University, Dhaka, Bangladesh. Previously worked as an Engineer in Security Operation Center at an InfoSec Consultant Farm (EIC). His research interests cover Machine Learning, Computer Vision, Natural Language Processing, Cloud Computing, and \& Cloud Security. 
\end{IEEEbiography}
\vspace{-1.2cm}
\begin{IEEEbiography}[{\includegraphics[width=1in,height=1.25in,clip,keepaspectratio]{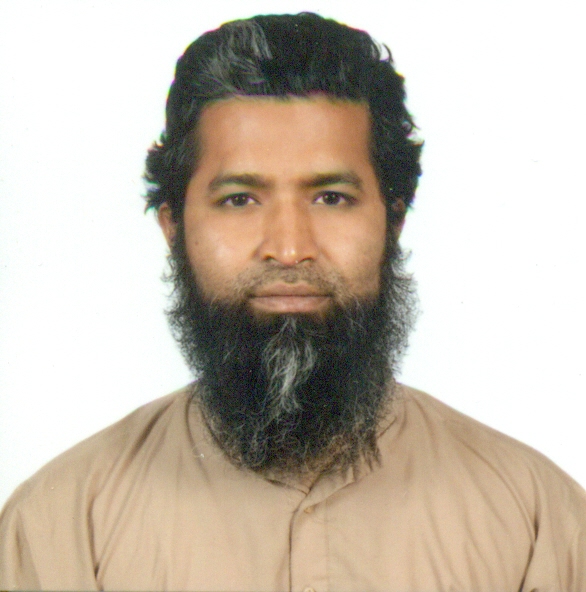}}]{A. B. M. Alim Al Islam} is working as Professor at Bangladesh University of Engineering and Technology, Dhaka, Bangladesh. He has received B.Sc. and M.Sc. Engineering in Computer Science and Engineering from Bangladesh University of Engineering and Technology, Bangladesh. He has also received a Ph.D. in Computer Science and Engineering, 2012 from the School of ECE, Purdue University, West Lafayette in the USA. His areas of research interests include Wireless Networks, Embedded Systems, Modeling \& Simulation, Computer Networks Security, and Reliability Analysis.
\end{IEEEbiography}

\vfill

\end{document}